\documentclass[aps,prb,twocolumn,showpacs,superscriptaddress]{revtex4}
\usepackage{graphicx}
\usepackage{bm}

\newcommand{\beq}{\begin{equation}}
\newcommand{\eeq}{\end{equation}}

\begin{document}

\title{\boldmath Anomalous Behavior of Thermodynamic Properties
                 of Strongly Correlated Fermi Systems }

\author{J.~W.~Clark}
\affiliation{McDonnell Center for the Space Sciences and
Department of Physics, Washington University,
St.~Louis, MO 63130, USA }
\author{V.~A.~Khodel}
\affiliation{Russian Research Center Kurchatov
Institute, Moscow, 123182, Russia}
\affiliation{McDonnell Center for the Space Sciences and
Department of Physics, Washington University,
St.~Louis, MO 63130, USA }
\author{M.~V.~Zverev}
\affiliation{Russian Research Center Kurchatov
Institute, Moscow, 123182, Russia}

\date{\today}

\begin{abstract}

Thermodynamic characteristics of  Fermi systems are investigated in the
vicinity of a phase
transition where the effective mass diverges and the single-particle
spectrum becomes flat.  It is demonstrated that at extremely low
temperatures $T$, the flattening
of the spectrum is reflected in non-Fermi-liquid behavior of the
inverse susceptibility $\chi^{-1}(T) \sim T^{\alpha}$ and the specific heat
 $C(T)/T\sim T^{-\alpha}$, with the critical index $\alpha=2/3$.
In the presence of an external static magnetic field
$H$, both these quantities are found to exhibit a scaling behavior, e.g.
$\chi^{-1}(T,H)=\chi^{-1}(T,0)+ T^{2/3}F(H/T)$, in agreement with
available experimental data.

\end{abstract}

\maketitle

Manifestations of non-Fermi-liquid (NFL) behavior observed
at extremely low temperatures $T$ in a number of strongly correlated
Fermi systems,\cite{godfrin1,godfrin2,saunders,krav,rez,coleman1,
stewart,takahashi,steglich,gegenwart}
notably $^3$He films and heavy-fermion metals, provide valuable clues to a
fundamental microscopic understanding of these systems.  This anomalous
behavior is commonly attributed to antiferromagnetic spin fluctuations, and
numerous strategies have been advanced (e.g.\ Ref.~\onlinecite{millis})
based on integrating out
all degrees of freedom except spin fluctuations.
However, the spin-fluctuation model (SFM) is unable to reproduce the
results of precise measurements of the inverse spin susceptibility,
which, on the ``metallic" side of the critical point, is found to obey
$\chi^{-1}(T)\sim T^{\alpha}$ with a critical
index $\alpha < 1$ (cf.\ Refs.~\onlinecite{godfrin1,gegenwart,coleman2}).
Difficulties are likewise encountered for the thermal expansion
data.\cite{grun}  The SFM also fails to
explain\cite{custers,coleman1,takahashi,gegenwart} the scaling
exhibited by $\chi(T,H)$ in external, sometimes tiny, static magnetic
fields $H$.  Moreover, the SFM falls short when confronted with
other generic features revealed by experiment, most notably a
divergence of the effective mass $M^*(\rho)$ at a critical density
$\rho=\rho_{\infty}$ and the emergence of NFL features even
before the critical point.\cite{godfrin1, saunders,krav,rez}

We are therefore compelled to explore a different strategy, in which NFL
anomalies are attributed to fermionic degrees of freedom and specifically
associated with flattening of the single-particle spectrum $\xi(p)$ in the
immediate vicinity of $\rho_{\infty}$.  Near this point, the FL formula
$\xi_{FL}(p)=p_F(p-p_F)/M^*(\rho)$ must be supplemented by terms nonlinear in
$p-p_F$, since $\xi_{FL}(p;\rho)$ vanishes identically at $\rho_{\infty}$.
Here we shall study consequences of the alteration of $\xi(p)$ on the
``metallic" side of the phase transition.  Importantly, as argued below,
the ratio of the damping $\gamma(\varepsilon)$ of single-particle excitations
to relevant energies $\varepsilon$ remains relatively small in this regime.
Accordingly, the Landau quasiparticle formalism still applies.

We focus our attention on the real part of the AC spin susceptibility
$\chi(T,\omega\to 0)$ given by the familiar FL formula \cite{lanl}
\beq
\chi(T,\rho) =
\chi_0(T,\rho)/[ 1-g_0(\rho) \Pi_0(T,\rho)]\,,
\label{chiform}
\eeq
where $g_0$ is the zeroth harmonic of the Landau spin-spin interaction
$\chi_0=-\mu^2_B\Pi_0$, and $\mu_B$ is the Bohr magneton. To be definite,
we treat the homogeneous three-dimensional (3D) case, with
\beq
\Pi_0(T)=
\int {dn(\xi(p))\over d\xi(p)} dv \equiv -{p^2_F\over \pi^2 T}
\int n(\xi)[1-n(\xi)]
{dp\over d\xi}d\xi \,,
\label{pio}
\eeq
where $n(\xi(p))=1/[1+\exp(\xi(p)/T)]^{-1}$ is the quasiparticle
momentum  distribution and $dv=2d^3p/(2\pi)^3$.  While Eqs.~(\ref{chiform})
and (\ref{pio}) are formally identical to familiar textbook formulas, the
function $\Pi_0(T)$
can differ profoundly from its FL realization due to
complicated character of the group velocity $d\xi(p)/dp$ in
the momentum region where $|\xi(p)|\simeq T$, which is
dominant in the integral (\ref{pio}).

To proceed, we invoke the well-known expression\cite{lanl}
\beq
{\partial \xi(p)\over \partial {\bf p}}={{\bf p}\over M}
+\int f({\bf p},{\bf p}_1)
{\partial n(\xi(p_1))\over \partial {\bf p}_1}dv_1 \,,
\label{lan}
\eeq
connecting the spectrum $\xi(p)$ and the momentum distribution $n(\xi)$
through the Landau interaction function $f({\bf p},{\bf p}_1)$.  At $T=0$,
Eq.~(\ref{lan}) implies that the effective mass $M^*$ is related to the
first harmonic $f_1(p_F,p_F)$ of the interaction function by
\beq
M/ M^*(\rho,T=0)=1-F^0_1(\rho)/3\equiv D(\rho)  \,,
\label{meff}
\eeq
in which
$F^0_1(\rho)=f_1(p_F,p_F)N_0$ with $N_0=p_FM/\pi^2$.  Thus $M^*(\rho,T=0)$
diverges at the critical density $\rho_{\infty}$, where
$D(\rho_{\infty})=0$, while at nonzero temperatures,
 $M^*(\rho_{\infty},T)$
already has a finite value.\cite{shag}
Its $T$ dependence is found by expanding relevant quantities on
both sides of Eq.~(\ref{lan}) in Taylor series, thereby obtaining
\beq
d\xi/dp\simeq p_F/M^*(\rho,T)+ v_2(p-p_F)^2/Mp_F
\label{gro}
\eeq
with
$$
{M\over M^*(T,\rho)}=D(\rho)+
{M\over 3p_F}\int\biggl[(f_1'p^2_F+2f_1p_F)(p-p_F)$$
\beq
+\left({1\over 2}f''_1p^2_F+2f'_1p_F+f_1\right)(p-p_F)^2\biggr]
{\partial n(\xi(p))\over \partial p}{dp\over \pi^2} \,.
\label{mt}
\eeq
Here $v_2=-p^3_FMf''_1/6\pi^2$, $f'_1= \left[df_1(p,p_F)/dp\right]_{p=p_F}$,
and $f''_1= \left[d^2f_1(p,p_F)/dp^2\right]_{p=p_F}$.
The term $v_1(\rho)(p-p_F)$, which provides the contribution
$v_1(\rho)(p-p_F)^2/2$ to $\xi(p)$, has been dropped in
writing Eq.~(\ref{gro}), since $v_1(\rho_{\infty})$ must vanish; otherwise
 the function $\xi(p,\rho_{\infty})$  has the same sign below and above
the Fermi surface, and the Landau state becomes unstable before $\rho$
reaches $\rho_{\infty}$.

Eq.~(\ref{mt}) can be simplified using particle-number
conservation, expressed approximately as
\beq
\int [p_F(p-p_F)+(p-p_F)^2]{\partial n(\xi(p))\over
\partial p}dp =0\,.
\label{pnc}
\eeq
Inserting this relation into Eq.~(\ref{mt}), we have
\beq
{M\over M^*(T,\rho)}=D(\rho)-v_2\int{(p-p_F)^2\over p^2_F}
{\partial n(\xi(p))\over \partial p} dp \,  .
\label{meft}
\eeq
In terms of the new variables $\tau=2TM/p^2_F$, $x=\xi(p)/T$ and
$y = (p-p_F)(v_2/3MTp_F)^{1/3}$, Eq.~(\ref{meft}) reduces to
$M/ M^*(T,\rho)=D(\rho)+\nu a\tau^{2/3}$, while
  Eq.(\ref{gro}) becomes
\beq
{d\xi(p,T,\rho)\over dp}={p_F\over M}\biggl[D(\rho)+\nu
\tau^{2/3}[a+
y^2)]\biggr] \  ,
\label{gro1}
\eeq
which is recast to a relation between $x$ and $y$,
\beq
 x=y\left[3\nu^{-1}D(\rho)\tau^{-2/3}+3a+y^2\right] \ .
\label{xy}
\eeq
where $\nu=(9v_2/ 4)^{1/3}$ and
\beq
     a =\int_{-\infty}^{\infty}y^2(x) e^x (1+e^x)^{-2} dx \,.
\label{kap}
\eeq
Having solved  Eqs.~(\ref{xy}) and (\ref{kap}),
the polarization operator $\Pi_0\equiv -N_0P_0$ is
straightforwardly evaluated to yield
\beq
\chi(T,\rho)=\mu^2_BN_0P_0(T,\rho)/[1+G_0(\rho)D(\rho)P_0(T,\rho)] \  ,
\label{chit}
\eeq
where $G_0=g_0p_FM^*(T=0)/\pi^2$,  and
\beq
P_0(T,\rho)=\int_{-\infty}^{\infty}{e^x (1+e^x)^{-2}
\over  [a(T,\rho)+y^2(x)]\nu\tau^{2/3}+D(\rho) } dx \ .
\label{pio2}
\eeq
In what follows we address systems without ferromagnetic ordering where
 $1+G_0>0$. In this case, $P_0(T,\rho)$ crucially depends on the ratio
 $T/|\rho-\rho_{\infty}|$. When $T$ drops to 0 while
$|\rho-\rho_{\infty}|$ is fixed, one obtains
$P_0(T,\rho)=
D^{-1}(\rho)$, and hence $\chi(T= 0,\rho\neq \rho_{\infty})=
\mu^2_BN_0/[D(\rho)(1+G_0)]\sim |\rho-\rho_{\infty}|^{-1}$. In the opposite
case, $D(\rho_{\infty})=0$ and
 calculations yield $M/M^*(T,\rho_{\infty})\simeq a(\rho_{\infty})
\nu\tau^{2/3}$, and $P_0(T,\rho_{\infty})=b(\rho_{\infty})
 \nu^{-1}\tau^{-2/3}$ with
$a(\rho_{\infty})=0.5$ and $b(\rho_{\infty})=1.2$, so that
\beq
\chi(T,\rho_{\infty})=\chi_0(T,\rho_{\infty})=1.2\mu^2_BN_0
\nu^{-1}\tau^{-2/3}\  .
\label{ch}
\eeq
  This implies that the critical index $\alpha$ specifying the
low-$T$ dependence of the inverse spin susceptibility
$\chi^{-1}(T,\rho_{\infty})$ is $2/3$.

It can be verified that  the corresponding 1D and 2D equations are
identical in form
to Eqs.~(\ref{xy}) and (\ref{kap}), derived here for the 3D case.

The above treatment can be extended to other thermodynamic
properties, e.g.\ the specific heat given by
\beq
C=
-{p_F^2\over \pi^2 }\int \xi n(\xi)[1-n(\xi)]{dp\over d\xi}
 {\partial\over \partial T}\biggl({\xi(T)\over T}\biggr) d\xi\,.
\label{spec}
\eeq
Manipulations similar to those applied to $\Pi_0(T,\rho)$ yield
\beq
{C(T,\rho)\over T} = N_0
\int_{-\infty}^{\infty}{x[x-2ay(x)]e^x (1+e^x)^{-2 }
\over [a+y^2(x)]\nu\tau^{2/3}+D(\rho) } dx \  .
\label{spet}
\eeq
 Thus we see that at $T\to 0$ the ratio $C(T)/T$ possesses the same
 NFL behavior as the spin susceptibility.

In electron systems of solids, as a rule there exist
several bands that cross the Fermi surface,
 the effective mass  diverging only in one or
two of them. Other bands still
contribute as Landau theory dictates providing  additional
 $T$-independent terms in $\chi(T)$ and $C(T)/T$ and modifying somewhat the
 prefactors in the singular parts. As a result, the behavior
$T^{-2/3}$ implied by Eq.(\ref{ch}) holds in
 dealing with the $T$-dependent excesses
$\Delta\chi(T,\rho_{\infty})$ and $\Delta C(T,\rho_{\infty})/T$.
One may  expect the same to be true for anisotropic
systems, e.g. for 2D liquid $^3$He, where the enhancement of $M^*$ is
associated
with the crystal lattice of the substrate. \cite{morishita}

\begin{figure}[t]
\includegraphics[width=0.9\linewidth,height=0.45\linewidth]{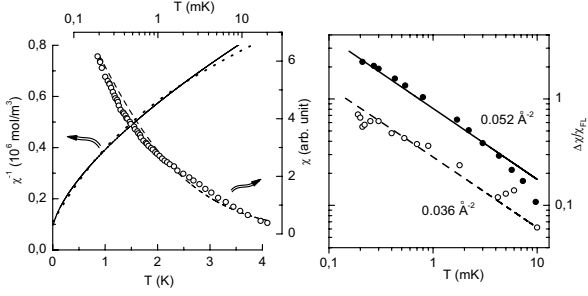}
\caption{Left panel, bottom-left axes: Inverse magnetic susceptibility
of YbRh(Si$_{0.95}$Ge$_{0.05})_2$ as a function of temperature.
Experimental data of Ref.~\onlinecite{steglich} are shown as short
dashes; the solid curve is the current prediction. Top-right axes:
temperature dependence of magnetic susceptibility of CeRu$_2$Si$_2$
in a magnetic field of 0.02 mT. Experimental data of
Ref.~\onlinecite{takahashi} are denoted by circles, and the
theoretical prediction by the dashed curve.
Right panel: Spin susceptibility excess,
divided by the FL contribution to the spin susceptibility,
evaluated for $^3$He films at two densities
(indicated near the curves).  Experimental data from
Ref.~\onlinecite{godfrin2} appear as solid and open circles, while
solid curves trace the predictions of the current theory at low $T$.
}
\label{fig:chi}
\end{figure}

To ascertain the relevance
of the basic result (\ref{ch}) to real strongly
correlated Fermi systems, we make use of  data
\cite{steglich,gegenwart,takahashi} for the heavy-fermion
metals YbRh(Si$_{0.95}$Ge$_{0.05}$)$_2$ and CeRu$_2$Si$_2$, where the
$T$-independent FL terms are rather small. These data, as seen from
the left panel of Fig.~\ref{fig:chi},  are well reproduced
by the proposed model.

In 2D liquid $^3$He the FL term in $\chi(T)$ is significant.
 Accordingly,
in the right panel of Fig.~\ref{fig:chi} we plot
 results for the excess $\Delta\chi(T)$
 at the two densities $\rho=0.036 A^{-2}$
 and $\rho=0.052A^{-2}$. In doing so, we suggest that   the flat
portion in the spectrum $\xi({\bf p})$
 forms in certain regions adjacent to
the Fermi line of $^3$He, and these regions produce contributions
$\Delta\chi(T)\sim T^{-2/3}$, while the remainder  generates only the FL terms.
We suggest that with increasing density  the latter region shrinks
and vanishes at the point where liquid $^3$He
solidifies. Our calculations ar compared with experimental data
 \cite{godfrin1}  in
the right panel of Fig.~\ref{fig:chi}.

Let us now examine the role of damping effects.  When evaluating
thermodynamic properties, characteristic energies $\varepsilon$
are of order of $T$.  We estimate the damping $\gamma(\varepsilon\sim T)$
with the help of the standard FLT formula
\beq
\gamma(\varepsilon\sim T)\sim  |\Gamma|^2\left[M^*(T)\right]^3T^2
\label{dam}
\eeq
containing the interaction amplitude $\Gamma$ and the effective mass
$M^*$, which according to Eq.~(\ref{gro1}), depends on $T$  as $T^{-2/3}$. It
is important to recognize that the customary replacement\cite{noz}
of $\Gamma$ by the bare interaction $V$, although
legitimate in ordinary Fermi liquids, is erroneous in the
limit of strong correlations.  The source of the error is
the huge enhancement of the density of states, which suppresses
$\Gamma$ and makes its magnitude quite insensitive to
the bare interaction.\cite{schuck}  To wit: summation of ladder
diagrams in the particle-hole channel gives $\Gamma=V/[1+VN(0)]$,
where $N(0) \sim p_FM^*$ is the density of states.  In the
limit $N(0)V\gg 1$, one obtains $\Gamma\sim N^{-1}(0)$.
Inserting this result into Eq.~(\ref{dam}),
we obtain
\beq
\gamma(\varepsilon\sim T)/ T\sim TM^*(T)/p^2_F\sim \tau^{1/3} \ll 1 \,,
\eeq
affirming the applicability of FLT theory to our problem.

Imposition of a static external magnetic field $H$ brings into play
a new dimensionless parameter $R=\mu_BH/T$ and opens another arena for
testing the model.  The function $n(\xi(p))$ entering Eq.~(\ref{lan})
is then replaced by $\left[n(\xi_+(p))+n(\xi_-(p)\right]/2$, where
$n\left(\xi_{\pm}(p)\right)=\left[1+\exp(\xi(p)/T \pm R/2) \right]^{-1}$.
In turn, $\xi(p)$ is determined from Eq.~(\ref{gro}), the effective
mass being the same for both spin directions
at sufficiently weak $H$.  Proceeding as before, we find
\beq
M/ M^*(T,H,\rho_{\infty})=\nu\tau^{2/3}a(R) \,,
\label{meh}
\eeq
where
 \beq
 a(R) ={1\over 2}\int y^2(x)
\biggl[{e^{x+R/2}\over (1+e^{x+R/2})^2}+{e^{x-R/2}\over(1+e^{x-R/2})^2}\biggr]
dx \,.
\label{kah}
\eeq

In the limit $T\to 0$ or equivalently $R\to \infty$, the solution
of this equation takes the analytic form
$M^*(T=0,H,\rho_{\infty})\sim H^{2/3}$,
confirming a result of Ref.~\onlinecite{shag}. Thus at sufficiently low
temperatures, imposition of static magnetic fields satisfying $\mu_BH>T$
renders
the effective mass $M^*(T,H,\rho_{\infty})$ finite, promoting the
recovery of the Landau theory.
\begin{figure}[t]
\includegraphics[width=0.8\linewidth,height=0.85\linewidth]{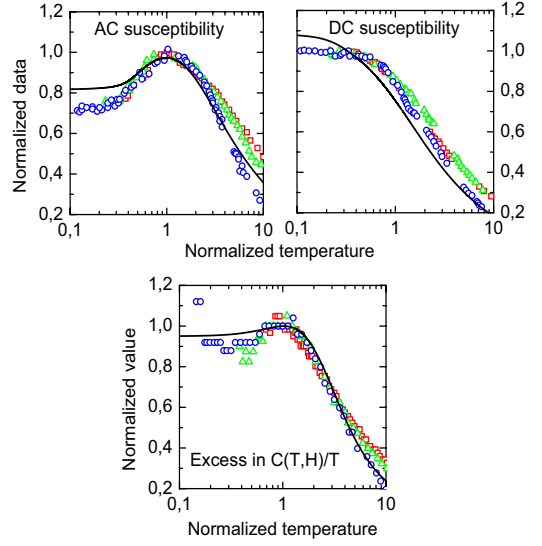}
\caption{Top panels: Normalized magnetic susceptibility
$\chi(T,H)/\chi(T_P)$ (top-left panel) and normalized
 magnetization ${\cal M}(T,H)/{\cal M}(T_P)$ (top-right panel) for
CeRu$_2$Si$_2$ in magnetic fields 0.20 mT (squares),
0.39 mT (triangles), and 0.94 mT (circles), plotted
against normalized temperature $T/T_P$ (Ref.~\onlinecite{takahashi}),
where $T_P$ is the temperature at peak susceptibility.
The solid curves trace the universal behavior predicted
by the present theory.
Bottom panel: The normalized ratio $C(T,H)T_M/C(T_M)T$
for YbRh(Si$_{0.95}$Ge$_{0.05}$)$_2$ in magnetic fields
0.05 T (squares), 0.1 T (triangles), 0.2 T (circles) versus
the normalized temperature $T/T_M$ (Ref.~\onlinecite{custers}),
where $T_M$ is the temperature at maximum ratio $C(T,H)/T$.
The solid curve shows the prediction of our theory.
}
\label{fig:sca}
\end{figure}
%
Along the same lines,  at the critical density $\rho_{\infty}$
we may establish that
 the magnetic moment and AC spin
susceptibility display a  scaling behavior, e.g.
\beq
\chi_{AC}(T,H,\rho_{\infty})=\mu_B^2N_0\nu^{-1}b(R)
\tau^{-2/ 3}  \  ,
\label{chi_r}
\eeq
where
\beq
b(R)={1\over 2}\int \left[ {e^{x+R/2} \over (1{+}e^{x+R/2})^2 }
+ {e^{x-R/2} \over (1{+}e^{x-R/2})^2 } \right]
  {dx \over a(R){+}y^2(x) }  \,.
\label{b_r}
\eeq

Turning to the specific heat, and following the same path
as taken above for $\chi(T,H)$ at the critical point, we are led
to the expression
\beq
C(T,H,\rho_{\infty})/T=N_0\nu^{-1}c(R) \tau^{-2/3}\,,
\label{heat_r}
\eeq
where
$$
c(R)={1\over 2}\int \biggl[ {[(x{+}R/2)^2{-}2(x{+}R/2)a(R)y(x)]e^{x+R/2}
\over (1+e^{x+R/2})^2}$$
\beq
+ {[(x{-}R/2)^2{-}(x{-}R/2)a(R)y(x)]e^{x-R/2}\over (1+e^{x-R/2})^2}\biggr]
{dx\over a(R){+}y^2(x) }  \,.
\label{c_r}
\eeq
A scaling behavior of this kind has been discussed by Coleman and
collaborators.\cite{coleman1,coleman2}

 For finite $H$,  the curve describing
$\chi_{AC}(T,H)$ acquires a maximum at some temperature $T_P$,  as does
the curve for $C(T,H)/T$,  at some temperature $T_M$.
This behavior is associated with    the
suppression of the divergent NFL terms $\sim T^{-2/3}$ in $\chi(T,H)$ and
$C(T,H)/T$
and recovery of the FL behavior  at static magnetic fields in which the
Zeeman energy splitting  $\mu_BH$ exceeds $T$.
 Following Ref.~\onlinecite{takahashi} we present in Fig.2
the results of numerical calculations of these quantities as functions
of the normalized temperatures $T/T_P$ and $T/T_M$ to demonstrate
that our model reproduces the experimental scaling behavior of the spin
susceptibility \cite{takahashi} of the heavy fermion metal CeRu$_2$Si$_2$
and that of the  specific heat \cite{custers} of the  heavy fermion compound
YbRh(Si$_{0.95}$Ge$_{0.05}$)$_2$,
 {\it without adjustable parameters}. It should be emphasized that
the curves
given in Fig.2 remain the same whether one is working with   heavy
fermion metals or 2D liquid $^3$He.
This universality, predicted by our model,  can be tested with the aid of  a new apparatus
\cite{japan} designed for measurements of thermodynamic properties of
 2D liquid $^3$He in static magnetic fields.

It is worth noting that, as seen from Eq.(\ref{pio}), on the metallic
side of the phase transition
the NFL behavior (\ref{ch})
holds in a  density  domain $\Delta\rho\sim \rho_{\infty}\tau^{2/3}$ adjacent to the critical point. On the other hand,
 as we plan to demonstrate  in our next article,
on the insulating side of the phase transition, the range of the
 domain where the  NFL term $\sim T^{-2/3}$ prevails
 is substantially larger, $\Delta\rho\sim \rho_{\infty}\tau^{1/3}$.

Let us compare our results with those available within the
antiferromagnetic scenario. \cite{millis}  First, the proposed
flattening mechanism
for NFL behavior adequately explains the low-$T$ data on the
spin susceptibility, producing $\chi^{-1}(T)\sim T^{\alpha}$
with $\alpha\simeq 2/3$ in the critical density region, while
 the SFM fails to
provide $\alpha<1$.  Second, the flattening mechanism
explains the scaling behavior  $\chi^{-1}(T)\sim T^{\alpha}F(H/T)$
of the spin susceptibility in static magnetic fields, whereas
such a scaling property does not arise in the SFM. Third, within
the model developed here, FL behavior can be recovered
at low $T$ close the critical point by imposing a tiny
magnetic field satisfying $\mu_BH>T$. In the SFM there is no
such provision for reinstating Fermi-liquid theory.

Finally, we discuss the relevance of ferromagnetic fluctuations to the NFL
 behavior, claimed in Refs.~\onlinecite{takahashi,gegenwart}.
Experimental data for most heavy-fermion systems show no
evidence of ferromagnetism.  The same is true for
 2D liquid $^3$He.\cite{godfrin1,saunders} This implies that
close to the  critical point,
 the respective Pomeranchuck stability condition \cite{lanl} $1+G_0(\rho)>0$
is not violated, in spite of the divergence of the effective mass at
$\rho=\rho_{\infty}$. The value of $G_0$
 can be estimated from the Landau sum rule\cite{lanl}
$\sum_L \left\{F_L/[1{+}F_L/(2L{+}1)]{+}G_L/[1{+}G_L/(2L{+}1)]\right\}=0$
by assuming all
the harmonics $F_L$ to be finite, apart from
$F_1{=}F^0_1M^*/M{\simeq}3M^*/M$.
The  Landau sum rule
is then satisfied provided $G_0{\simeq}-0.75$.
This value
of $G_0$ gives rise to a marked enhancement
of the Sommerfeld ratio in the case $\mu_BH>T$, to a around $12$ that
 changes little if
the zeroth harmonic $F_0$ is taken into account.
Even so, such an enhancement does not in itself ensure
the relevance of ferromagnetic fluctuations to the NFL
behavior. Indeed, dense
3D liquid $^3$He, in which $G_0\simeq -0.8$, shows no deviations
from FLT predictions as $T\to 0$.

In summary, we have studied the impact of flattening of the
single-particle spectrum on the spin susceptibility
$\chi(T)$ and the specific heat $C(T)$ of strongly correlated
Fermi systems.  When the density approaches
the critical value at which effective mass becomes infinite,
Fermi-liquid theory progressively fails as the deviant components
in $\chi(T)$ and $C(T)$ grow to dominance.  We have explicated
the implied critical behavior of these components and derived
a universal relation between them.  Finally, our analysis has
revealed the scaling behavior of $\chi(T,H)$ and $C(T,H)/T$ in
the presence of an external magnetic field $H$.  Numerical
results based on this theoretical picture are in agreement
with experimental data on 2D liquid $^3$He and several
heavy-fermion metals.

We thank L.~P.~Gor'kov, G.~Kotliar, E.~Krotscheck,
V.~R.~Shaginyan,
and V.~M.~Yakovenko for valuable discussions.  This research
was supported by NSF Grant PHY-0140316, by the
McDonnell Center for the Space Sciences, and by
Grant NS-1885.2003.2 from the
  Russian Ministry of Education and Science.

\end{document}